\documentstyle[aps,prl,multicol,subeqnarray,cite,enumerate,verbatim]{revtex}
\begin{document}
\newcommand{\dis}{\displaystyle}
\newcommand{\dml}{\rm dim ~ }
\newcommand{\krl}{\rm ker ~ ~ }
\newcommand{\expon}{\rm e}
\newcommand{\id}{ 1 \hspace{-2.85pt} {\rm I} \hspace{2.5mm}}
\newcommand{\beq}{\begin{equation}}
\newcommand{\eeq}{\end{equation}}
\newcommand{\bseq}{\begin{subeqnarray}}
\newcommand{\eseq}{\end{subeqnarray}}
\renewcommand{\thefootnote}{\fnsymbol{footnote}}
\newcommand{\tr}{\mbox{\rm tr}}
 \newcommand{\g}{{\bf g}}
\newcommand{\rivpt}{\rule{.4pt}{4pt}}
\newcommand{\leftmk}{\makebox[8cm]{\hrulefill\rivpt}}
\newcommand{\rightmk}{\hspace{9cm}\rivpt\raisebox{1ex}{\makebox[8cm]{\hrulefill}}}

\draft 

\title{A unified approach for exactly solvable potentials in quantum mechanics using shift operators}
\author{Mo-Lin Ge$^a$, L. C. Kwek$^b$,  Yong
Liu$^{abc}$ {\footnote{Email address:phyliuy@nus.edu.sg}}, C. H.
Oh$^b$ {\footnote{E-mail address: phyohch@leonis.nus.edu.sg }}and
Xiang-Bin Wang $^b${\footnote{E-mail address: phywangx@nus.edu.sg
}} }
\address{$^a$ Theoretical Physics Division, Nankai Institute of
Mathematics, Tianjin 300071, People's Republic of China}
\address{$^b$Department of Physics, Faculty of Science, \\
National University of Singapore, Lower Kent Ridge, \\
Singapore 119260, Republic of Singapore. }
\address{$^c$Laboratory of Numerical Study for Heliospheric Physics (LHP), \\
Chinese Academy of Science, \\ P.O. Box 8701, Beijing 100080, P.R. China. }
\maketitle


\begin{abstract}
We present a unified approach for solving and classifying exactly
solvable potentials.  Our unified approach encompasses many
well-known exactly solvable potentials. Moreover, the new approach
can be used to search systematically for a new class of solvable
potentials.
\end{abstract}

\bigskip
 \begin{multicols}{2}
\section{Introduction}

Exactly solvable potentials are extensively applied in the
investigation of many physical systems in condensed matter,
nuclear physics, quantum optics and solid state physics. Typical
examples within this large class of potentials in non-relativistic
quantum mechanics include the Morse potential\cite{morse}, the
P\"{o}schl-Teller potential and the Coulomb potential.  In
particular, the Morse potential has provided a reasonably suitable
approximation for studying anharmonicity and bond dissociation of
diatomic molecules at low vibrational levels. Recently, coherent
states \cite{benedict}, $q$-deformation \cite{cooper95, gupta} and
the phase space and transport properties of coupled Morse
potentials \cite{zembekov} have also been investigated. The
P\"{o}schl-Teller potential on the other hand has always been
explored for its short-range properties. Its generalized coherent
states \cite{crawford}, nonlinear properties \cite{liu, quesne}
and supersymmetric extension\cite{diaz} have also been extensively
studied. In recent years, under the guise of supersymmetric
quantum mechanics, the idea of shaped invariant potentials have
also been mooted \cite{gendenstein, dutt86, cooper88, khare,
dabrowska}. In particular, a large class of such potential is the
Natanzon class \cite{natanzon, fukui,levai,foot1}.

Algebraic methods\cite{kamran, celeghini} exploiting the
underlying Lie symmetry and its associated algebras have been
widely used to study many of these exactly solvable potentials,
for instance, among many others, Darboux transformation,
Infeld-Hull factorization, Mielnick factorization, SUSY quantum
mechanics, inverse scattering theory \cite{jens} and intertwining
technique \cite{diaz}. Moreover, by exploiting an underlying
$q$-deformed quantum algebraic symmetry,
Spiridonov\cite{spiridonov} has shown how a finite difference
differential equation can generate in the limiting cases a set of
exactly solvable potentials. Recently, algebraic method based on
nonlinear algebras \cite{liu, quesne,gupta} has also been used to
analyze these solvable potentials. The main purpose of this paper
is to develop a new unified approach to obtain transition
operators for a large class of exactly solvable potentials.

In section \ref{method}, we briefly sketch the main ideas behind
the diagonalization process. We then sketch  the main idea behind
our method of getting shift operators from Hamiltonians with
arbitrary solvable potentials in section \ref{shift}. In section
\ref{solvable}, we illustrate our method by showing how most of
the exactly solvable potentials can be obtained through the
method.  Finally in section \ref{discuss}, we summarize our
technique and discuss some subtle points associated with it.

\section{Method}\label{method}
Suppose we have a physical system with Hamiltonian $ H$,
generalized coordinate operator, $ Q$, and generalized momentum
operator, $ P$, satisfying the commutation relations\cite{nieto}
\bseq \label{comm} \lbrack  H,  Q \rbrack & = &  Q \Theta_1(H,
I_i, c_i) +  P \Pi_1 (H, I_i, c_i) , \\ \lbrack  H,  P \rbrack & =
&  Q \Theta_2(H, I_i, c_i)  +  P \Pi_2 (H, I_i, c_i), \eseq where
${ I_i}, \; i \in {\cal S}$ for some arbitrary index set, ${\cal
S}$, are some invariants of motion of the system and $c_i, \; i
\in {\cal S}$ are some constants. Denoting
\begin{equation}
([{ H},\; { Q}],\; [{ H},\;  { P}]):= [{ H}, \; ({ Q}, \; { P})],
\end{equation}
we find that we can recast the previous commutation equations more
succinctly as
\begin{equation}
([{ H},\; { Q}],\; [{ H},\;  { P}]) U= [{ H},\; ({ Q}, \; { P})
U],
\end{equation}
where $U$ is a $2 \times 2$ matrix so that the commutation
relations in eq(\ref{comm}) can be rewritten as
\begin{equation}\label{coef1}
[{ H},\;  ({ Q}, \; { P}) ]=({ Q}, \; { P}) \left(
\begin{array}{cc}
\Theta_1 & \Theta_2\\ \Pi_1 & \Pi_2
\end{array}
\right) \equiv ({ Q}, \; { P}) M.
\end{equation}
By diagonalizing the matrix, $M$,  on the right hand side of
eq(\ref{coef1}), we see that $$ [{ H},\; ({ Q}, \; { P})U ]=({ Q},
\; { P})UU^{-1} M U $$ giving $$ [{ H},\; ({ Q}, \; { P}) U ]=({
Q}, \; { P}) U \left(
\begin{array}{cc}
E_1 & 0\\ 0& E_2
\end{array}
\right), $$ where $E_1$ and $E_2$ are the eigenvalues of $M$. By
defining $({ Q}, \; { P})U$ as $(S_1, S_2)$, we see that the
operators $S_1$ and $S_2$ acts as lowering and raising operators.

In the specific case of the one dimensional harmonic oscillator
\cite{raab,cooper}, the Hamiltonian is given by $ {
H}=\frac{1}{2}({ x}^2+{ p}^2)$ where ${ x}$ is the coordinate
operator and ${ p}=-i\frac{d}{dx}$ is the momentum operator with
commutator given by $[{ x},\; { p}]=i$. Moreover, it is obvious
from the Hamiltonian that the following commutation relations hold
$ [{ H}, \;{ x}]=-i { p}, ~ [{ H}, \; { p}]=i { x}$. Identifying $
Q$ and $ P$ as $ x$ and $ p$, we see that
\begin{equation}\label{coef}
[{ H},\;  ({ x}, \; { p}) ]=({ x}, \; { p}) \left(
\begin{array}{cc}
0 & i\\ -i & 0
\end{array}
\right).
\end{equation}
Diagonalizing the matrix on the right hand side of eq(\ref{coef})
gives $$ [{ H},\; ({ x}, \; { p})U ]=({ x}, \; { p})UU^{-1} \left(
\begin{array}{cc}
0 & i\\ -i & 0
\end{array}
\right)U $$ so that \begin{eqnarray*} [{ H},\;  ({ x}, \; { p})
\frac{1}{\sqrt{2}}\left(
\begin{array}{cc}
1&1\\ i&-i
\end{array}
\right) ] & = & \\ & & \mbox{\hspace{-2.5cm}}({ x}, \; { p})
\frac{1}{\sqrt{2}} \left(
\begin{array}{cc}
1&1\\ i&-i
\end{array}
\right) \left(
\begin{array}{cc}
-1 & 0\\ 0&1
\end{array}
\right). \end{eqnarray*} Thus,
\begin{equation}
[{ H},\;  ({ a}, \; { a^\dagger}) ]=({ a}, \; { a^\dagger}) \left(
\begin{array}{cc}
-1 & 0\\ 0 & 1
\end{array}
\right),
\end{equation}
and we can therefore identify $\dis { a}\equiv
\frac{1}{\sqrt{2}}(x + i p)$ and ${ a^\dagger} \equiv
\frac{1}{\sqrt{2}}(x - i p)$ as the natural shift operators
arising from the diagonalization process. The harmonic oscillator
is the simplest example. In general, it is possible to replace the
coordinate operator ${ x}$ and momentum operator ${ p}$ by
generalized ``coordinate" operator ${ Q}$ and ``momentum" operator
${ P}$ which are functions of the operators ${ x}, \; { p}$, ${
H}$ and $I_i, i \in {\cal S}$. Physically, the energy spectrum
will no longer be uniformly spaced since the eigenvalues, $E_1$
and $E_2$, associated with the raising and lowering operators are
in functions of $H$ and the amount through which these operators
raise or lower depend very much on the energy level as well. This
is the main source of nonlinearity in the algebra. Moreover, we
note that the coefficient matrix, $M$, in eq(\ref{coef1}) as well
as the corresponding transformation matrix, $U$, are generally
functions of the Hamiltonian $H$, the invariants of motion $I_i$
and some constants $c_i$.

\section{Shift Operators}\label{shift}
We next see how the shift operators can be obtained for
Hamiltonians with arbitrary solvable potentials. This method can
also be regarded as a means of searching systematically for the
solvable potentials. To this end, we begin with the following
Hamiltonian
\begin{equation}
{ H}=X(x) \frac{d^2}{dx^2}+V(x).
\end{equation}
where $X(x)$ is an arbitrary function of position and $V(x)$ is an
arbitrary potential. Define
\begin{equation}
{ P}({ x}, { p})=Y(x) \frac{d}{dx}+Z(x)
\end{equation}
with arbitrary functions $Y(x)$ and $Z(x)$ to be determined. It is
easy to show that the commutation relation between $ H$ and $ P$
is
\begin{eqnarray}
 [{ H}, \; { P}] & = &(2 X(x) Y'(x)-X'(x) Y(x))
\frac{d^2}{dx^2}\nonumber \\ & & + X(x) (Y''(x)+2 Z'(x))
\frac{d}{dx}+X(x) Z''(x)\nonumber \\ & & \mbox{\hspace{1cm}}-Y(x)
V'(x).
\end{eqnarray} Since $X(x)$, $Y(x)$ and $Z(x)$ are arbitrary
functions, it is possible to set
\begin{eqnarray}
\label{e1} X(x) ( Y''(x)+2 Z'(x) )&=&\alpha Y(x)\\ \label{e2} 2
X(x) Y'(x)-X'(x) Y(x)&=&(\beta { Q}(x) +\gamma)X(x),
\end{eqnarray}
where $\alpha$, $\beta$ and $\gamma$ are arbitrary constants and $
Q(x)$ is the generalized ``coordinate" to be determined, so that
\begin{eqnarray} [{ H}, \; { P}]& = & (\beta { Q}(x)
+\gamma) { H} +\alpha { P}-(\beta { Q}(x)+\gamma) V(x)\nonumber
\\& & -\alpha Z(x) -Y(x) V'(x)+X(x) Z''(x).
\end{eqnarray} Furthermore, if we impose for simplicity \begin{eqnarray}
& &  X(x) Z''(x)-(\beta { Q}(x)+\gamma) V(x)-\alpha Z(x)\nonumber
\\ & & \mbox{\hspace{1cm}}
-Y(x) V'(x)={ Q}(x), \end{eqnarray} and solved for the operator $
Q(x)$ as
\begin{eqnarray}
\label{e3} { Q}(x)& = & \frac{1}{1+\beta V(x)}( X(x) Z''(x)-\gamma
V(x)\nonumber \\ & & \mbox{\hspace{2cm}}-\alpha Z(x)-Y(x) V'(x) ),
\end{eqnarray}
we find
\begin{equation}
[{ H}, \; { P}]={ Q}(x) (\beta { H} +1) +\alpha { P}+\gamma { H}.
\end{equation}
Concomitantly, we can also work out explicitly the expression for
the commutator of ${ H}$ with ${ Q}(x)$ giving \beq [{ H}, \; {
Q}(x)]=2 X(x) { Q}'(x)\frac{d}{dx} + X(x) { Q}''(x). \eeq Finally,
one can set
\begin{equation}
\label{e4} X(x) { Q}'(x)=\lambda Y(x),
\end{equation}
so that $$ [{ H}, \; { Q}(x)]=2 \lambda { P}-2 \lambda Z(x) + X(x)
{ Q}''(x). $$ If the further constraint
\begin{equation}
\label{e5} -2 \lambda Z(x) + X(x) { Q}''(x)=\nu { Q}(x)+\tau,
\end{equation}
is imposed, we arrive at
\begin{equation}
[{ H}, \; { Q}(x)]=2 \lambda { P}+\nu { Q}(x)+\tau.
\end{equation} Here, $\lambda$, $\nu$ and $\tau$ are constants to be
determined. Note that the quantities $\tau$ and $\gamma { H}$ can
be absorbed in certain cases by a redefinition of the operators ${
Q}(x)$ and ${ P}$ so that we finally obtain a semi-closed algebra.
In the process, the parameters $\alpha, \; \beta, \; \gamma, \;
\lambda , \; \nu, \; \tau$ are not arbitrary but must satisfy some
constraints. It is also interesting to note that in the above
construction of the shift operators, we are not constrained by
commutation relation between ${ Q}(x)$ and ${ P}$. Moreover, we do
not need to require the operators ${ H}, { Q}$ and ${ P}$ form a
closed algebra, although for most of the well known exactly
solvable Hamiltonians, these operators always inevitably form a
closed algebra. However, to obtain the ground state, we need to
require that the Hamiltonian factorizes according to $ S_1 S_2
\sim F(H, I_i, c_i)$ for some arbitrary function, $F$.

In the construction, we should bear in mind that $Y(x), \; Z(x), {
Q}(x)$ and $\alpha, \; \beta, \; \gamma, \; \lambda , \; \nu, \;
\tau$ are all determined by $X(x)$ and $V(x)$ when we construct
the shift operators for a given Hamiltonian. Conversely, if we
proceed with a systematic search for exactly solvable potentials
based on this method, we can have as many degrees of freedom as
there are free parameters and functions, namely $X(x), \; Y(x), \;
Z(x), { Q}(x)$ and $\alpha, \; \beta, \; \gamma, \; \lambda, \;
\nu, \; \tau$.

\section{Some solvable potentials}\label{solvable}
We can perform a systematic search for a class of exactly solvable
potentials by solving the equations
(\ref{e1},\ref{e2},\ref{e3},\ref{e4},\ref{e5}). Furthermore, the
method is tantamount to a construction the corresponding shift
operators so that it is possible to obtain the energy spectrum and
the underlying algebraic formulation if necessary. Indeed, most of
well known solvable potentials can be reproduced using the method
by an appropriate choice of functions and parameters.

{\bf Case} 1. $X(x)=-1, \;\; Y(x)=1$.

On substituting into the equations
(\ref{e1},\ref{e2},\ref{e3},\ref{e4},\ref{e5}), consistency requires
the imposition of the following constraints \begin{eqnarray*} 
Q^\prime(x) & = & -\lambda, \\ - \beta Q(x) - \gamma & = & 0, \\ -
2 Z^\prime (x) & = & \alpha, \\ - 2 \lambda Z(x) - Q^{\prime
\prime} (x) & = & \nu Q(x) + \tau, \\ V^\prime (x) + (\beta Q(x) +
\gamma )V(x) & = & - Z^{\prime \prime}(x) - \alpha Z(x) - Q(x).
\end{eqnarray*}
It is easy to solve these constrained equations to obtain $$ {
Q}(x)=-\lambda x+c_1, \;\; Z(x)=-\frac{\alpha}{2}x+c_2, $$ $$
V(x)=\frac{1}{2} (\lambda+\frac{\alpha^2}{2}) x^2 -(\alpha
c_2+c_1) x+c_3 $$ and $\beta=0, \; \gamma=0,\; \nu=-\alpha$ and
$\tau=c_1 \nu-2 \lambda c_2$ with the integral constants $c_i,
i=1, 2, 3$. This is the harmonic oscillator.

By substituting the solution of the constrained equations into
eq(\ref{coef1}), we get \begin{eqnarray*}
\lbrack H, Q \rbrack & = & - \alpha Q + 2 \lambda P - (2 \lambda
c_2 + \nu c_1), \\ \lbrack H, P \rbrack & = & Q + \alpha P.
\end{eqnarray*}
A simple redefinition of the operators $Q$ and $P$ using $Q =
\tilde{Q} + f$ and $P = \tilde{P} + g$ where $\dis f= - \alpha
\frac{2 \lambda c_2 + \nu c_1}{ \alpha^2 + 2 \lambda}$ and $\dis g
= \frac{2 \lambda c_2 + \nu c_1}{\alpha^2 + 2 \lambda}$ so that
the operators $\tilde{Q}$ and $\tilde{P}$ take the form \begin{eqnarray*}
\tilde{Q} & = & - \lambda x + c_1 + \alpha \frac{2 \lambda c_2 +
\nu c_1 }{\alpha^2 + 2 \lambda} \\ \tilde{P} & = & \frac{d}{dx}-
\frac{\alpha}{2} x + c_2 - \frac{2 \lambda c_2 + \nu c_1}{\alpha^2
+ 2 \lambda} \end{eqnarray*}
immediately give the equations
\begin{eqnarray*}
\lbrack H, \tilde{Q} \rbrack & = & - \alpha \tilde{Q} + 2 \lambda
\tilde{P}, \\ \lbrack H, \tilde{P} \rbrack & = & \tilde{Q} +
\alpha \tilde{P},
\end{eqnarray*}
or more succinctly as $$ [H, (\tilde Q, \tilde P)] = (\tilde Q,
\tilde P) \left( \begin{array}{cc} - \alpha & 1 \\ 2 \lambda &
\alpha \end{array} \right)$$ which can be easily diagonalized to
give the shift operators $S_1$ and $S_2$ which satisfies the
commutation relation $$ [H, (S_1, S_2)] = (S_1, S_2) \left(
\begin{array}{cc} -\sqrt{\alpha^2 + 2 \lambda} & 0 \\ 0 &
\sqrt{\alpha^2 + 2 \lambda} \end{array} \right)$$ with
\begin{eqnarray*} S_1 & = & \frac{1}{\alpha - \sqrt{\alpha^2 + 2
\lambda}}\tilde{Q} + \tilde P \\ 
S_2 & = & \frac{1}{\alpha + \sqrt{\alpha^2 + 2 \lambda}}\tilde{Q}
+ \tilde
P \end{eqnarray*} and \beq \label{case1eq1}
[S_1, S_2]  =  - \sqrt{\alpha^2 + 2 \lambda}\eeq It is clear from
eq(\ref{case1eq1}) that the shift operators $S_1$ and $S_2$ are
really the usual lowering and raising operators resepctively for
the harmonic oscillators ( simply set $\dis A = \frac{-S_1}{
\left( \alpha^2 + 2 \lambda \right)^{1/4}}$ and $\dis A^\dagger =
\frac{S_2}{ \left( \alpha^2 + 2 \lambda \right)^{1/4}}$). Having
constructed the shift operators, the ground state eigenvalue and
eigenvector can be solved by the action of the lowering operator
on a lowest weight state using the equations \beq H \psi_0(x) =
E_0 \psi_0(x), \ \  S_1 \psi(x) = 0. \eeq  The results are easily
found to be
\begin{eqnarray} \psi_0 & = & B_3 \exp \left( \frac{1}{2} B_1 x^2
+ B_2 x \right)
\\ E_0 & = & \frac{1}{2} \sqrt{\alpha^2 + 2 \lambda} + c_3 -
\frac{(c_1 + c_2 \alpha)^2 }{\alpha^2 + 2 \lambda},
\end{eqnarray}
where $B_i, i=1,2,3$ are constants of integration.

{\bf Case} 2. $X(x)=-1, \;\; Y(x)=x$.

As in Case 1, a simple substitution into the equations
(\ref{e1},\ref{e2},\ref{e3},\ref{e4},\ref{e5}) requires for
consistency the following constraints \begin{eqnarray*} 
Q^\prime(x) & = & - \lambda x, \\ - \beta Q(x) - \gamma & = & -2,
\\ - 2 Z^\prime (x) & = & \alpha x, \\ - 2 \lambda Z(x) - Q^{\prime
\prime} (x) & = & \nu Q(x) + \tau, \\ x V^\prime (x) + (\beta Q(x)
+ \gamma )  V(x) & = & - Z^{\prime \prime}(x) - \alpha Z(x) -
Q(x).
\end{eqnarray*}  By solving the equations, we obtain $$ {
Q}(x)=-\frac{\lambda}{2}x^2+c_1, \;\; Z(x)=-
\frac{\alpha}{4}x^2+c_2, $$ $$V(x)=
\frac{1}{8}(\lambda+\frac{\alpha}{2})^2 x^2+\frac{c_3}{x^2}+
\frac{1}{2}(\frac{\alpha}{2}-\alpha c_2-c_1) $$ with $\beta=0, \;
\gamma=2, \; \nu=-\alpha$ and $\tau=\lambda (1-2 c_2)+\alpha c_1$.
This gives us the radial harmonic oscillator potential.  The rest
of the details are similar to Case 1, but we will provide some of
the crucial results.

The shift operators $S_1$ and $S_2$ can be found through a similar
method with the explicit form of the operators given by
\begin{eqnarray*}
S_1 = \tilde Q + (\alpha - \sqrt{\alpha^2 + 2 \lambda}) \tilde P
\\
S_2 = \tilde Q + (\alpha + \sqrt{\alpha^2 + 2 \lambda}) \tilde P
\end{eqnarray*}
as before but with the redefined operators $\tilde Q$ and $\tilde
P$ defined by
\begin{eqnarray*}
\tilde Q  & = & -\frac{\lambda}{2} x^2 + c_1 + \frac{4
\lambda}{\alpha^2 + 2 \lambda} H \nonumber \\ & &
\mbox{\hspace{2cm}} - \frac{\alpha }{\alpha^2 + 2 \lambda}
(\lambda + \alpha c_1 - 2 \lambda c_2 ) \\ \tilde P  & = & x
\frac{d}{dx}  - \frac{\alpha }{4} x^2 + c_2 + \frac{2
\alpha}{\alpha^2 + 2 \lambda} H \nonumber \\ & &
\mbox{\hspace{2cm}} + \frac{1 }{\alpha^2 + 2 \lambda} (\lambda +
\alpha c_1 - 2 \lambda c_2 ).
\end{eqnarray*} Moreover, the following commutation relation,
similar to Case 1, is
also satisfied $$ [H, (S_1, S_2)] = (S_1, S_2) \left(
\begin{array}{cc} -\sqrt{\alpha^2 + 2 \lambda} & 0 \\ 0 &
\sqrt{\alpha^2 + 2 \lambda} \end{array} \right).$$
 Note that these operators depend on $H$ so that
the eigenvalues associated with the raising and lowering operators
depend on the energy level it is acting on. Thus the tower of
states is no longer equally spaced.
The commutation relation for the shift operators is \beq 
[S_1, S_2]  =  \frac{8 \lambda}{\sqrt{\alpha^2 + 2 \lambda}} (2 H
 + c_1 + c_2 \alpha - \frac{\alpha }{2}). \eeq The ground state
can be shown to be \beq \psi_0  =  B_4 x^{-(B_2 E_0 + B_3)} \exp(-
\frac{B_1}{2} x^2) \eeq with energy given implicitly by
\beq 
(B_2 E_0 + B_3)^2 +(B_2 E_0 + B_3) - c_3 =0. \eeq

{\bf Case} 3. $X(x)=-1, \;\; Y(x)=a e^{c x}+b e^{-c x}$.

A similar substitution into equations
(\ref{e1},\ref{e2},\ref{e3},\ref{e4},\ref{e5}) and solving the
resulting consistency equations give the solution $$ {
Q}(x)=-\frac{\lambda}{c} (a e^{c x}-b e^{-c x}) + c_1,$$ $$
Z(x)=-\frac{\alpha+ c^2}{2 c} (a e^{c x}+b e^{-c x})+ c_2,$$ $$
V(x)=c_3 (a e^{c x}+b e^{-c x})^{-2}+ \frac{1}{4 c^2}[(\alpha +
c^2)^2 + 2 \lambda ] $$ and $\beta=-\frac{2}{\lambda}c^2,
\;\gamma=\frac{2 c^2 }{\lambda} c_1, \; \nu= -\alpha - 2 c^2,\;
\tau= - 2 \lambda c_2 - \nu c_1$. This is the generalized
P\"{o}schl-Teller potential\cite{cod}. Moreover, we also have the
the relation $c_1 + \alpha c_2= 0$.

The shift operators can be found similarly using the technique
described in Case 1 and the commutation relations for the shift
operators, $S_1$ and $S_2$, are
\begin{eqnarray}
\lbrack H, S_1 \rbrack & = &  S_1 \left(-c^2 - \sqrt{(\alpha +
c^2)^2 + 2 \lambda - 4 c^2 H } \right) \\ \lbrack H, S_2\rbrack &
= & S_2 \left(-c^2 + \sqrt{(\alpha + c^2)^2 + 2 \lambda - 4 c^2 H
} \right).
\end{eqnarray}
The explicit form of the shift operators are given by
\begin{eqnarray*}
S_1 & = & -\tilde Q \frac{1}{2 \lambda}((\alpha + c^2) -
\sqrt{(\alpha + c^2)^2 - 4 c^2 H + 2\lambda }) + \tilde P
\\ S_2 & = &- \tilde Q \frac{1}{2 \lambda}((\alpha +
c^2) + \sqrt{(\alpha + c^2)^2 -4 c^2 H + 2\lambda }) + \tilde P
\end{eqnarray*}
where
\begin{eqnarray*}
\tilde Q & = & Q + \frac{\gamma}{\beta}= Q - c_1, \\ \tilde P & =
& P - c_2.
\end{eqnarray*} with commutation relation $$ [\tilde Q, \tilde P]
= \lambda \left( a e^{c x} + b e^{- c x} \right)^2,$$ and
\beq 
\lbrack S_1, S_2 \rbrack = - 8 a b c \sqrt{(\frac{\alpha + c^2}{2 c})^2 +
\frac{\lambda}{2 c^2} - H}
\eeq

The ground state is given by \beq \psi_0 = c_4 (a e^{c x} + b e^{-
c x})^{-
\sqrt{(\frac{\alpha + c^2}{2 c})^2 + \frac{\lambda}{2 c^2} - E_0} }\eeq and the
ground
state energy $E_0$ can be determined through the equation \begin{eqnarray}
& & \left( (\frac{\alpha + c^2}{2 c})^2 + \frac{\lambda}{2 c^2}- E_0
\right)\nonumber \\
& & \mbox{\hspace{1cm}} + c \sqrt{(\frac{\alpha + c^2}{2 c})^2 + \frac{\lambda}{2
c^2}- E_0 } + \frac{c_4}{4 a b}= 0. \end{eqnarray}

{\bf Case} 4. $X(x)=-1, \;\; Y(x)=a \sin(k x) +b \cos(kx)$.

For this case, we get $$ { Q}(x)=\frac{\lambda}{k}(a \cos(k x) -b
\sin(kx)) + c_1,$$ $$ Z(x)=\frac{\alpha - k^2}{2 k}(a \cos(k x)- b
\sin(kx))+ c_2,$$ $$ V(x)=c_3(a \sin(k x) +b
\cos(kx))^{-2}-\frac{(k^2 - \alpha)^2 + 2 \lambda}{4k^2}$$ and $
\beta=\frac{2 k^2}{\lambda}, \;\gamma=- \frac{2}{\lambda} c_1
k^2,\;\nu=2 k^2- \alpha, \; \tau = c_1(\alpha - 2 k^2) - 2 \lambda
c_2$. This is the familiar P\"{o}schl-Teller potential.

The shift operators are given by
\begin{eqnarray*}
\lbrack H, S_1 \rbrack & = & S_1 (k^2 - \sqrt{(\alpha - k^2)^2 + 2
\lambda + 4 k^2 H }) \\ \lbrack H, S_2 \rbrack & = & S_2 (k^2 +
\sqrt{(\alpha - k^2)^2 + 2 \lambda + 4 k^2 H })
\end{eqnarray*}
with the operators $S_1$ and $S_2$ defined by
\begin{eqnarray*}
S_1 & = & (a \sin (k x) + b \cos (k x))\frac{d}{dx} \nonumber \\ &
&\mbox{\hspace{-1cm}} - \frac{1}{2 k} (a \cos (k x) - b \sin (k
x)) \sqrt{(\alpha - k^2)^2 + 2 \lambda + 4 k^2 H } \\ S_2 & = & (a
\sin (k x) + b \cos (k x))\frac{d}{dx} \nonumber \\ & &
\mbox{\hspace{-1cm}} + \frac{1}{2 k} (a \cos (k x) - b \sin (k x))
\sqrt{(\alpha - k^2)^2 + 2 \lambda + 4 k^2 H }
\end{eqnarray*}
and the commutator given by \beq [S_1, S_2] = - 2k(a^2 + b^2)
\sqrt{H + \frac{\lambda}{2 k^2} + (\frac{\alpha - k^2}{2 k})^2}.
\eeq The ground state, $\psi_0$, can be solved and found to be
given by the formula \beq \psi_0 = c_4 (a \sin (k x) + b \cos (k
x))^{\frac{1}{2 k^2} \sqrt{(\alpha - k^2)^2 + 2 \lambda + 4 k^2
E_0 }} \eeq where $E_0$ is the ground state energy determined
through the equation \begin{eqnarray} \frac{4 k^2}{a^2 + b^2} c_3
+ 2 k^2 \sqrt{(\alpha - k^2)^2 + 2 \lambda + 4 k^2 E_0 } & &
\nonumber \\ - \left( (\alpha - k^2)^2 + 2 \lambda + 4 k^2 E_0
\right) & = & 0.
\end{eqnarray}

{\bf Case} 5. $X(x)=-x, \;\; Y(x)=x$.

A similar substitution of above trial functions yields $$ {
Q}(x)=-\lambda x+c_1, \; Z(x)=-\frac{\alpha}{2}x+c_2,$$ $$V(x)=
\frac{1}{2}(\lambda + \frac{\alpha^2}{2})x+\frac{c_3}{x}-(c_1 +
\alpha c_2) $$ and $\beta=0, \; \gamma=1,\; \nu=-\alpha, \;
\tau=\alpha c_1 -2 \lambda c_2$. This is the case of Coulomb
potential for the hydrogen atom.

The shift operators, $S_1$ and $S_2$, satisfy the commutation
relations \begin{eqnarray*} \lbrack H, S_1 \rbrack & = & -
\sqrt{\alpha^2 + 2 \lambda} S_1 \\ \lbrack H, S_2 \rbrack & = &
\sqrt{\alpha^2 + 2 \lambda} S_2,
\end{eqnarray*}
in the same manner as the case for the harmonic oscillator in Case
1 and the radial harmonic oscillator in Case 2 but with the
operators defined as $$ (S_1, S_2)= (\tilde Q, \tilde P)U $$ where
\begin{eqnarray*}
\tilde Q & = & - \lambda x + \frac{2 \lambda}{\alpha^2 + 2
\lambda} H + 2 \lambda \frac{c_1 + \alpha c_2 }{\alpha^2 + 2
\lambda}, \\ \tilde P & = &  x \frac{d}{dx} - \frac{\alpha}{2} x +
\frac{\alpha}{\alpha^2 + 2 \lambda} H + \alpha \frac{c_1 + \alpha
c_2 }{\alpha^2 + 2 \lambda},
\end{eqnarray*}
and the matrix $U$ given by $$ U =\left( \begin{array}{cc}
\frac{1}{\alpha - \sqrt{\alpha^2 + 2 \lambda}} & \frac{1}{\alpha +
\sqrt{\alpha^2 + 2 \lambda} } \\ 1 & 1
\end{array}\right). $$
It is instructive to note that the commutation relation for the
shift operators can be computed succinctly as $$ [S_1, S_2] = -
\frac{2}{\sqrt{\alpha^2 + 2 \lambda}} (H + c_1 + \alpha c_2).$$ A
direct computation using $S_1 \psi_0 = 0$ yields the ground state
as \beq \psi_0 = c_4 x^{- \frac{c_1 + \alpha c_2 +
E_0}{\sqrt{\alpha^2 + 2 \lambda}}} e^{\frac{\sqrt{\alpha^2 + 2
\lambda}}{2} x}, \eeq where the ground state energy $E_0$ is given
by \beq E_0 = \frac{\sqrt{\alpha^2 + 2 \lambda}}{2} (1 \pm \sqrt{1
+ 4 c_3}) - c_1 - \alpha c_2, \eeq choosing appropriately the
correct minimal value.

{\bf Case} 6. $X(x)=- e^{c x}, \;\; Y(x)=1$.

As a final case, we look the above choice, so that upon
substitution into the appropriate equations gives $$ {
Q}(x)=\frac{\lambda}{c} e^{-cx}+c_1, \; Z(x)=\frac{\alpha}{2
c}e^{-c x} + c_2, $$ $$V(x)=\frac{2 \lambda+\alpha^2 }{4 c^2}e^{-c
x}+c_3 e^{c x}+\frac{\alpha}{2 c}( 2 c_2 -c ) + \frac{c_1}{c} $$
and $ \beta=0, \; \gamma=-c, \;  \nu=-\alpha, \; \tau=\alpha c_1 -
\lambda(c+ 2 c_2) $. We note that this is simply the case of Morse
potential. As in Case 5, the shift operators, $S_1$ and $S_2$,
obey the relations
\begin{eqnarray*} \lbrack H, S_1 \rbrack & = & -
\sqrt{\alpha^2 + 2 \lambda} S_1 \\ \lbrack H, S_2 \rbrack & = &
\sqrt{\alpha^2 + 2 \lambda} S_2,
\end{eqnarray*} with the shift operators defined by
$$ (S_1, S_2)= (\tilde Q, \tilde P)U $$ where
\begin{eqnarray*}
\tilde Q & = &  \frac{\lambda}{c} e^{- c x} - \frac{2 \lambda c
}{\alpha^2 + 2 \lambda} H , \\ \tilde P & = &   \frac{d}{dx} +
\frac{\alpha}{2 c} e^{- c x} - \frac{\alpha c}{\alpha^2 + 2
\lambda} H,
\end{eqnarray*}
and the matrix $U$ given by $$ U =\left( \begin{array}{cc}
\frac{1}{\alpha - \sqrt{\alpha^2 + 2 \lambda}} & \frac{1}{\alpha +
\sqrt{\alpha^2 + 2 \lambda} } \\ 1 & 1
\end{array}\right) $$ as before.
The commutation relation between the shift operators, $S_1$ and
$S_2$, now reads \beq [S_1, S_2] = -\frac{2 c^2}{\sqrt{\alpha^2 +
2 \lambda} } (H + \frac{2 c_1 + \alpha (c + 2 c_2)}{2 c}). \eeq
Moreover, a straightforward calculation yields the ground state as
\begin{eqnarray} \psi_0 & = & c_4 \exp {\biggl\{} {-\frac{\sqrt{\alpha^2 + 2 \lambda}}{2 c^2}
e^{- c x} } \nonumber \\& & \mbox{\hspace{-1cm}} + \frac{\alpha(c
+ 2 c_2) + c \sqrt{\alpha^2 + 2 \lambda }  + 2 c_1 - 2 c E_0}{2
\sqrt{\alpha^2 + 2 \lambda}} x {\biggl\}},\end{eqnarray} and the
ground state energy, $E_0$, is determined by \begin{eqnarray} E_0
& = & \frac{1}{2c}{\Bigl\{}2 c_1 + \alpha (c + 2 c_2) \nonumber\\
& & + c \sqrt{\alpha^2 + 2 \lambda} \pm \sqrt{2 c_3 (\alpha^2 + 2
\lambda)} {\Bigl\}},
\end{eqnarray} with an appropriate choice of sign.

Finally, we would like to comment briefly on the last two cases.
Normally, for radial Coulomb problem and Morse oscillator, we deal
with the following Schr\"{o}dinger equations $$
(-\frac{d^2}{dr^2}+\frac{l(l+1)}{r^2}-\frac{2
Z}{r}+\frac{Z^2}{n^2})\psi_{v, l}(r)=0 $$ and $$
(-\frac{d^2}{dy^2}+l^2 (1-e^{-y})^2-l^2)\psi_{v,
l}=-(l-v-\frac{1}{2})^2\psi_{v, l} $$ respectively. However, it is
always possible to transform the above equations into  more
convenient forms as $$ (-\rho
\frac{d^2}{d\rho^2}+\frac{l(l+1)}{\rho}+\rho)\psi_{v, l}=2 n
\psi_{v,l} $$ with $\rho=\frac{Z}{n}r$ and $$ (-e^x
\frac{d^2}{dx^2}+(l-v-\frac{1}{2})^2 e^{x}+e^{-x})\psi_{v, l}=2
l\psi_{v, l}. $$ To get a deeper insight into this transformation,
we note that the Hamiltonian concerned can be rewritten as $$ {
H}=-\frac{d^2}{dx^2}+R(x)+\frac{1}{T(x)}, $$ where $T(x)$ and
$R(x)$ are suitably chosen functions. The eigenvalue problem for
stationary eigenstates takes the form \beq
(-\frac{d^2}{dx^2}+R(x)+\frac{1}{T(x)})\psi(x)=E\psi(x). \eeq
Multiplying by the function $T(x)$, we see that the same
eigenvalue problem now appear as \beq (-T(x) \frac{d^2}{dx^2}+T(x)
(R(x)-E))\psi(x)=-\psi(x). \eeq Hence, the original eigenvalue
problem is now transformed into an equivalent eigenvalue problem
with the Hamiltonian \beq { H}'=-T(x) \frac{d^2}{dx^2}+T(x)
(R(x)-E). \eeq  It is precisely with the new form that the
potential can be solved more easily. For Coulomb and Morse
potentials, we have therefore used the modified Hamiltonian.

\section{Discussion and Conclusion} \label{discuss}

It is known that the local behavior of most solvable potentials
reduces  either to the harmonic oscillator or the
P\"{o}schl-Teller potential\cite{wehr}. This fact is further
confirmed by the commutation relations between the Hamiltonian and
the shift operators in section \ref{solvable}.  A quick glance at
the various cases clearly shows that the general form for the
commutation relations can be classified into two major categories,
those belonging to the harmonic oscillator (Case 1, Case 2, Case 5
and Case 6), corresponding to $\beta=0$, and those belonging to
the P\"{o}schl-Teller potential (Case 3 and Case 4), corresponding
to $\beta \neq 0$. Thus, algebraically, we should expect similar
behavior for potentials belonging to the same class.

The inherent nonlinearity in the algebraic structure of the shift
operators, in which $F_i(H) S_i = S_i G_i(H), i = 1,2$, should
have a deep physical connections to the study of exactly solvable
potentials in supersymmetric theory \cite{cooper,levai2}.
Moreover, the technique presented can be systematically analyzed
to encompass some of these supersymmetric potentials.  Indeed, in
section \ref{solvable}, we have not entirely exhausted all
possibilities and provided a comprehensive list. However, it is in
principle possible to do so. By exploring more complicated
functions $X(x)$, $Y(x)$ and $Z(x)$, it is expected that more
complicated solvable potentials may arise.

Recently, there has been much interest generated regarding the
computation of Franck-Condon factors for anharmonic
oscillators\cite{herzberg,muller}. In particular the Morse and
P\"{o}schl-Teller potentials do not in general provide a simple
analytical expression for the Franck-Condon factors\cite{carvajal,
iachello}. Our technique should be able to cast new light on the
exact computation of the Franck-Condon factors for such
potentials. More work in this direction will be reported
elsewhere. Finally, we note that the method can be extended easily
to many-body potentials. This work was supported primarily by NUS
Research Grant No. RP3982713. In addition, we would also like to
thank the referee for his invaluable comments.

\end{multicols}

\begin{thebibliography}{99}
\bibitem{foot1} Note that
there also exist non-shaped invariant Natanzon potentials
\cite{cooper, levai}.
\end{thebibliography}
\end{document}